\documentclass[conference]{IEEEtran}
\IEEEoverridecommandlockouts
\usepackage{mathrsfs}
\usepackage{dsfont}
\usepackage{amsfonts}
\usepackage{changepage}
\usepackage{bm}
\usepackage{booktabs}
\usepackage{multirow}
\usepackage{makecell}

\usepackage{textcomp,booktabs,threeparttable, amssymb}
\usepackage[usenames,dvipsnames]{color}
\usepackage{colortbl}
\definecolor{mygray}{gray}{.9}
\definecolor{mypink}{RGB}{150, 250, 190}
\definecolor{mycyan}{RGB}{255, 230, 190}

\usepackage[pdftex]{graphicx}
\graphicspath{{../pdf/}{../jpeg/}}
\DeclareGraphicsExtensions{.pdf,.jpeg,.png}

\usepackage[cmex10]{amsmath}
\usepackage{array}
\usepackage{mdwmath}
\usepackage{eqparbox}
\usepackage{url}
\usepackage{hyperref}
\usepackage[hyphenbreaks]{breakurl}

\usepackage{cite}
\usepackage{amsmath,amssymb,amsfonts}
\usepackage{algorithmic}
\usepackage{graphicx}
\usepackage{textcomp}
\usepackage{xcolor}
\def\BibTeX{{\rm B\kern-.05em{\sc i\kern-.025em b}\kern-.08em
    T\kern-.1667em\lower.7ex\hbox{E}\kern-.125emX}}
\begin{document}

\title{Nonparametric End-to-End Probabilistic Forecasting of Distributed Generation Outputs Considering Missing Data Imputation\\
\thanks{The authors gratefully acknowledge the funding support by Guangzhou Power Supply Bureau of Guangdong Power Grid Co. LTD (GDKJXM20222332). Corresponding author: Zichao Meng, e-mail: mzc20@mails.tsinghua.edu.cn}
}

\author{\IEEEauthorblockN{1\textsuperscript{st} Minghui Chen}
\IEEEauthorblockA{\textit{Guangzhou Power Supply Bureau} \\
\textit{Guangdong Power Grid Co., Ltd.}\\
Guangzhou, China \\
chenminghui1985@qq.com}
\and
\IEEEauthorblockN{2\textsuperscript{nd} Zichao Meng*}
\IEEEauthorblockA{\textit{Tsinghua-Berkeley Shenzhen Institute (TBSI)} \\
\textit{Tsinghua University}\\
Shenzhen, China \\
mzc20@mails.tsinghua.edu.cn}
\and
\IEEEauthorblockN{3\textsuperscript{rd} Yanping Liu}
\IEEEauthorblockA{\textit{Guangzhou Power Supply Bureau} \\
\textit{Guangdong Power Grid Co., Ltd.}\\
Guangzhou, China \\
liuliujx\textunderscore{1018@163.com}}
\and
\IEEEauthorblockN{4\textsuperscript{th} Longbo Luo}
\IEEEauthorblockA{\textit{Guangzhou Power Supply Bureau} \\
\textit{Guangdong Power Grid Co., Ltd.}\\
Guangzhou, China \\
104267600@qq.com}
\and
\IEEEauthorblockN{5\textsuperscript{th} Ye Guo}
\IEEEauthorblockA{\textit{Tsinghua-Berkeley Shenzhen Institute (TBSI)} \\
\textit{Tsinghua University}\\
Shenzhen, China \\
guo-ye@sz.tsinghua.edu.cn}
\and
\IEEEauthorblockN{6\textsuperscript{th} Kang Wang}
\IEEEauthorblockA{\textit{Tsinghua-Berkeley Shenzhen Institute (TBSI)} \\
\textit{Tsinghua University}\\
Shenzhen, China \\
wang.kang@sz.tsinghua.edu.cn}
}

\maketitle

\begin{abstract}

  In this paper, we introduce a nonparametric end-to-end method for probabilistic forecasting of distributed renewable generation outputs while including missing data imputation. Firstly, we employ a nonparametric probabilistic forecast model utilizing the long short-term memory (LSTM) network to model the probability distributions of distributed renewable generations' outputs. Secondly, we design an end-to-end training process that includes missing data imputation through iterative imputation and iterative loss-based training procedures. This two-step modeling approach effectively combines the strengths of the nonparametric method with the end-to-end approach. Consequently, our approach demonstrates exceptional capabilities in probabilistic forecasting for the outputs of distributed renewable generations while effectively handling missing values. Simulation results confirm the superior performance of our approach compared to existing alternatives.
\end{abstract}

\begin{IEEEkeywords}
Distributed renewable generation, end-to-end probabilistic forecasting, missing data imputation
\end{IEEEkeywords}

\section{Introduction}
Distributed generation refers to small-scale energy generation facilities located in different places, such as solar panels, wind turbines, etc., which directly supply power to end-users \cite{dugan2002distributed}. Compared with traditional centralized generation, distributed generation systems have advantages such as decentralization, flexibility, and environmental friendliness, but they also face a series of challenges, especially the impact of their stochastic nature. The global energy transition has accelerated the development of distributed generation systems. The limitations of traditional energy sources, environmental protection pressures, and the demand for energy security have gradually increased the proportion of renewable energy in the energy mix. Due to its renewable nature, decentralized layout, and flexibility, distributed generation has gradually become one of the main trends in energy transition.

Distributed generation systems exhibit high levels of randomness, mainly manifested in several aspects. 1) Weather Conditions: Renewable energy sources such as solar and wind energy are significantly influenced by weather conditions, leading to uncertainty in output power. For example, solar power generation is affected by factors like cloud cover and changes in light intensity \cite{rocchetta2015risk}. 2) Load Fluctuations: The energy demand of end-users experiences significant fluctuations due to factors such as industrial production and household electricity consumption, resulting in random changes in load \cite{driesen2006distributed}. 3) Equipment Reliability: The operational status and maintenance conditions of equipment in distributed generation systems also affect the randomness of generation. Factors such as equipment failures and maintenance cycles are sources of randomness \cite{jikeng2011reliability}.

To effectively characterize the stochastic nature of distributed generation systems for better prediction, management, and optimization of system operations, various techniques are employed, which are listed as follows. 1) Probability and Statistical Methods \cite{ruiz2012probabilistic}: Describing the distribution patterns of random variables through probability distribution functions, statistical analysis, etc., such as normal distribution, Poisson distribution, etc. 2) Time Series Analysis \cite{athari2017time}: Analyzing historical data to identify patterns and trends in time series, for instance, Autoregressive Moving Average (ARMA), Autoregressive Integrated Moving Average (ARIMA), etc. 3) Monte Carlo Simulation \cite{el2006investigating}: Conducting multiple simulations using random sampling methods to obtain the distribution characteristics and probability distribution of random variables, suitable for simulating complex systems. 4) Artificial Intelligence and Machine Learning \cite{jumaa2021optimal}: Utilizing methods such as neural networks, support vector machines, random forests, etc., through big data analysis and learning, to predict and optimize the stochastic characteristics of distributed generation systems.

Characterizing the stochastic nature of distributed generation systems is crucial for achieving stable operation and optimal management. However, randomness leads to uncertainty and complexity in system operation. Specifically, randomness makes it difficult to accurately predict the future operational state of the system, increasing the risks in system planning and operation \cite{agalgaonkar2015stochastic}. Additionally, the volatility caused by randomness may affect the stability and reliability of the system, requiring effective management and control strategies. Furthermore, influenced by various factors such as weather changes and equipment failures, accurate characterization of the stochastic nature of distributed generation requires effective missing data handling techniques. Missing values are a common problem in distributed generation data analysis, which may result from sensor failures, data collection errors, etc. The existence of missing values may lead to incomplete data, thereby affecting the accuracy of subsequent data analysis and prediction models \cite{sundararajan2020evaluation}. Therefore, effective missing data imputation techniques are crucial to infer missing values from existing data, making the dataset more complete.

In this context, the development of missing data imputation techniques has become a key research direction in the field of distributed generation. The development of this technology aims to solve data quality problems and improve the accuracy of data analysis and prediction. For missing values, interpolation methods may be used to infer missing values based on the correlation of existing data \cite{ang2020research}, or machine learning algorithms may be employed to predict and fill missing values, thereby achieving data completeness \cite{maruf2018locating}. These technologies may involve big data processing and distributed computing since distributed generation systems typically generate large amounts of data, requiring the use of big data frameworks and distributed processing techniques for effective outlier cleaning and missing data imputation. Currently, existing methods mainly proceed through independent two-phase processes, first imputing missing values and then characterizing the stochastic nature of distributed generation based on the completed dataset. This makes it difficult to control the imputation errors in the stage of stochastic nature characterization, thereby reducing the accuracy of stochastic nature characterization.

To address these issues, this paper proposes an end-to-end approach for probabilistic forecasting of distributed generation considering missing data imputation, aiming to mitigate the negative impact of missing data imputation on the characterization of distributed generation stochasticity. This approach achieves a more accurate characterization of distributed generation stochasticity considering data missingness.

\section{Problem Description and Metrics}

\subsection{Problem Description}

Probabilistic forecasting of distributed energy resources with missing data aims to determine the probability distribution of distributed energy outputs by addressing missing values through imputation. The historical observations of the power outputs of the distributed energy up to time step $t$ with missing data can be represented as:

\begin{equation}
\boldsymbol{\tilde{X}}_t = [\cdots, x_{t-5}, \slash, \slash, x_{t-2}, x_{t-1}, \slash],
\label{miss-data}
\end{equation}
where $x_t$ denotes the observed distributed energy output at time $t$, and "$\slash$" indicates missing data at the corresponding time step. Before forecasting, this incomplete vector $\boldsymbol{\tilde{X}}_t$ is imputed to create a complete sequence:
\begin{equation}
\boldsymbol{X}_t = [\cdots, x_{t-5}, \hat{x}_{t-4}, \hat{x}_{t-3}, x_{t-2}, x_{t-1}, \hat{x}_{t}],
\label{imputation}
\end{equation}
where $\hat{x}_{*}$ denotes an imputed value for the missing distributed energy data.

After imputation, the probability distribution of outputs of distributed energy can be estimated using a probabilistic forecast model, which maps historical observations to quantiles of the forecasting target:

\begin{equation}
x_{t+l|t}^{\alpha} = F(\boldsymbol{X}_t, \alpha; \theta),
\label{nonp-model}
\end{equation}
where $x_{t+l|t}^{\alpha}$ represents the $\alpha$-th quantile prediction of distributed energy output at lead time $l$ (here we consider characterize the stochastic properties in the future), $Q$ is the set of quantiles of interest, and $F(\cdot; \theta)$ is the probabilistic forecast model parameterized by $\theta$. If there are no missing data in $\boldsymbol{X}_t$, the probabilistic forecasting of distributed energy resources is implemented directly using the observed historical data without imputation.

Quantile regression, a nonparametric method, is employed for modeling the probabilistic forecast model $F(\cdot; \theta)$, due to its flexibility and lack of reliance on specific distribution assumptions. The parameters $\theta$ of the probabilistic forecast model are optimized by minimizing the pinball loss $L$:
\begin{equation}
L = \sum_{\alpha \in Q}\alpha \cdot \text{max}(0, x_{t+l} - x_{t+l|t}^{\alpha}) + (1-\alpha) \cdot \text{max}(0, x_{t+l|t}^{\alpha} - x_{t+l}),
\label{pinball-loss}
\end{equation}
where $x_{t+l}$ represents the observed distributed energy output at time $t+l$, and $x_{t+l|t}^{\alpha}$ is the forecasted quantile. This loss function ensures that the forecast captures both overestimation and underestimation uncertainties, corresponding to the $\alpha$ quantile.

\subsection{Evaluation Metrics}

The forecasted quantiles are evaluated within the probabilistic forecast evaluation framework outlined in \cite{Pinson-WE-2007}.

Reliability assessment of a probabilistic forecast model involves measuring the average deviations between expected and observed frequencies below the forecasted quantiles:
\begin{equation}
\overline{R_l}=\frac{1}{I}\sum_{i=1}^I|\alpha_{i}-\frac{1}{N}\sum_{n=1}^NH({x}_{n+l|n}^{\alpha_i}-x_{n+l})|\text{,}
\label{a_l}
\end{equation}
where $N$ represents the sample count in the testing dataset. The expected frequency, denoted by the nominal proportion $\alpha_{i}$, spans from $5\%$ to $95\%$ ($I=19$) in $5\%$ increments. ${x}_{n+l|n}^{\alpha_i}$ indicates the $\alpha_i$-th forecasted quantile with an observation of $x_{n+l}$, while $H(x)$ denotes the unit step function.

Sharpness assessment quantifies the average width of prediction intervals (PI) across different levels (1-$\alpha_{i}$):
\begin{equation}
  {\overline{S_l}}=\frac{1}{I\cdot N}\sum_{i=1}^I\sum_{n=1}^N({x}_{n+l|n}^{1-\alpha_{i}/2}-{x}_{n+l|n}^{\alpha_{i}/2})\text{.}
  \label{d_l}
\end{equation}

The skill score integrates reliability and sharpness, and the average skill score over $N$ time spots is computed as:
\begin{equation}
  \overline{Sk_l}=\frac{1}{N}\sum_{n=1}^N\sum_{i=1}^I \{[H({x}_{n+l|n}^{\alpha_{i}}-x_{n+l})-\alpha_{i}](x_{n+l}-{x}_{n+l|n}^{\alpha_{i}})\}\text{.}
  \label{s_l}
\end{equation}

\section{The Proposed Method}
\label{sec-3}

\subsection {Framework}

\begin{figure}[t]
  \begin{center}
  \includegraphics[width=2.5in]{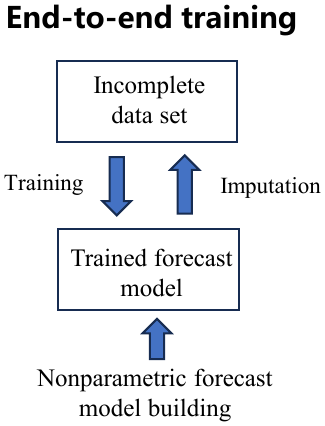}\\
  \caption{Nonparametric end-to-end framework for probabilistic forecasting of distributed energy resources considering missing data imputation.}\label{overview}
  \end{center}
\end{figure}

The framework of the method for probabilistic forecasting of distributed energy resources considering missing data imputation is depicted in Fig.~\ref{overview}. We first build a nonparametric probabilistic forecast model to depict the probabilistic forecasting of distributed energy resources. Next, we introduce the end-to-end training for the probabilistic forecast model considering missing data inputation.

\subsection {Building Nonparametric Probabilistic Forecast Model}

Considering the effectiveness of modeling temporal relations in RNN-based algorithms, we use LSTM \cite{greff2016lstm} to build the deep-learning-based model in this paper. Specifically, the LSTMs adopted here share the same deep residual structure as the one demonstrated in \cite{Hu-TNLS-2020} to alleviate the gradient vanishing problem in deep neural networks. For example, when considering a lag interval denoted as $\delta$, the input for the LSTM can be represented as $\boldsymbol X_t=[x_{t-\delta+1}\text{, }\cdots\text{, }x_t]$. In the recurrent calculation steps of the LSTM, ranging from time step $\kappa=t-\delta+1$ to $\kappa=t$, the operations unfold as follows
\begin{figure}[!t]
  \begin{center}
  \includegraphics[width=3.3in]{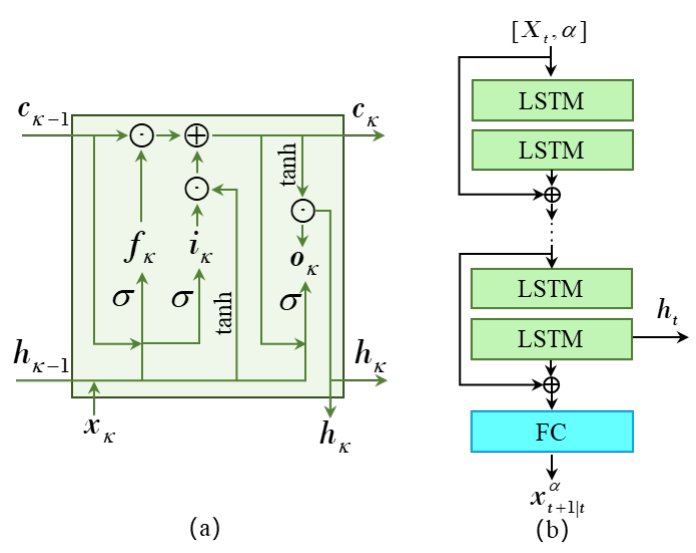}\\
  \caption{LSTM layer and model. (a) is a demonstration of the calculation steps in one LSTM layer. (b) is the structure of the nonparametric probabilistic forecast model.}\label{forecast-model}
  \end{center}
\end{figure}

\begin{figure*}[!t]
  \begin{center}
  \includegraphics[width=5.8in]{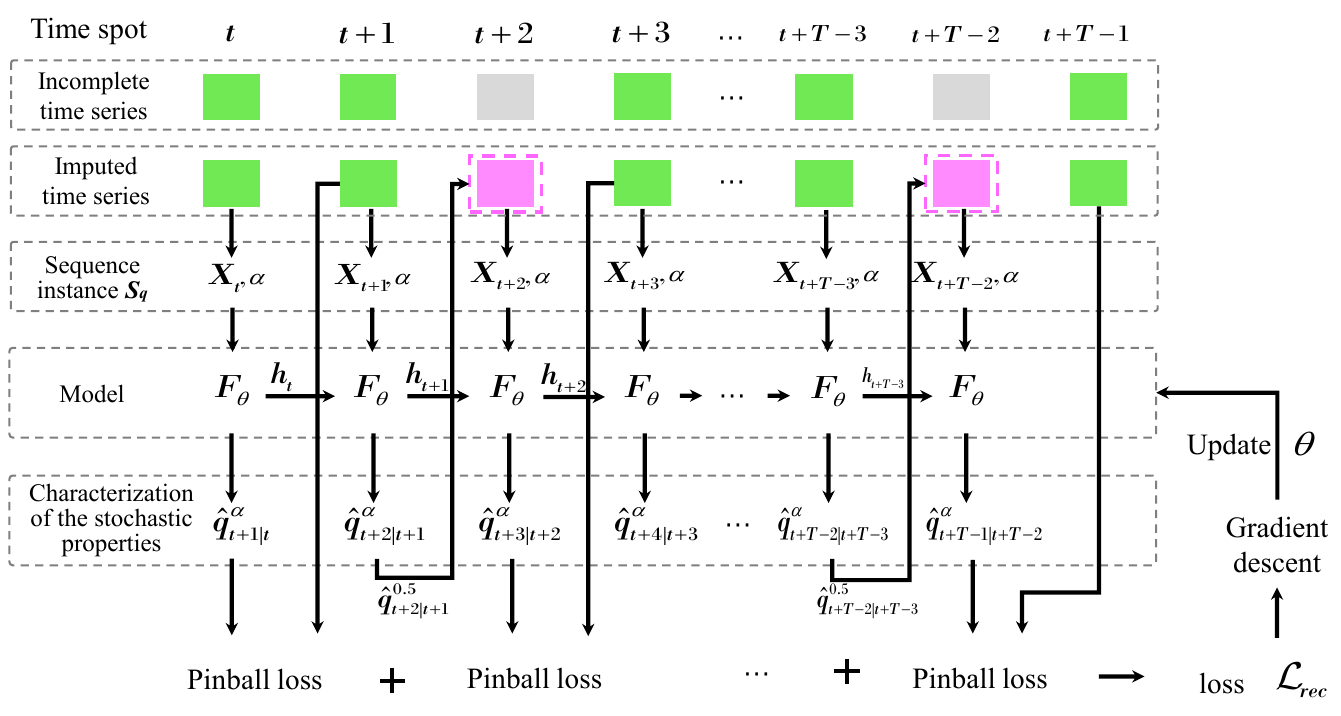}\\
  \vspace{-0.2cm}
  \caption{End-to-end training considering missing data imputation.}\label{end-to-end-trian}
  \end{center}
\end{figure*}
\begin{equation}
{i}_\kappa=\sigma({w}_{ix}\cdot{x}_\kappa+{w}_{ih}\cdot{h}_{\kappa-1}+{w}_{ic}\cdot{c}_{\kappa-1}+{b}_{i})\text{,}\label{lstm-b}
\end{equation}
\begin{equation}
{f}_\kappa=\sigma({w}_{fx}\cdot{x}_\kappa+{w}_{fh}\cdot{h}_{\kappa-1}+{w}_{fc}\cdot{c}_{\kappa-1}+{b}_{f})\text{,}
\end{equation}
\begin{equation}
{c}_\kappa={f}_\kappa\odot{c}_{\kappa-1}+{i}_\kappa\odot\text{tanh}({w}_{cx}\cdot{x}_\kappa+{w}_{ch}\cdot{h}_{\kappa-1}+{b}_{c})\text{,}\label{lstm-c}
\end{equation}
\begin{equation}
{o}_\kappa=\sigma({w}_{ox}\cdot{x}_\kappa+{w}_{oh}\cdot{h}_{\kappa-1}+{w}_{oc}\cdot{c}_{\kappa-1}+{b}_{o})\text{,}
\end{equation}
\begin{equation}
  {h}_\kappa={o}_\kappa\odot\text{tanh}({c}_\kappa)\text{,}\label{lstm-e}
  \end{equation}
where ${i}_\kappa$, ${f}_\kappa$ and ${o}_\kappa$ represent input gate, forget gate and output gate, respectively. The vectors ${h}_\kappa$ and ${c}_\kappa$ represent cell state and hidden state, respectively, with both having the same size in terms of the number of elements. The tensors ${w}$ with different subscripts denote weights, and ${b}$ with different subscripts denote biases. The symbol $\odot$ signifies the element-wise product. The recurrent calculation steps (\ref{lstm-b})-(\ref{lstm-e}) collectively form a single LSTM layer, as illustrated in Fig.~\ref{forecast-model}(a).

As shown in Fig.~\ref{forecast-model}(b), we incorporate the nominal proportion $\alpha$ in the input, and formulate the forecast model with many LSTM layers with residual connections and one fully connected (FC) layer to finally generate quantiles of forecasts. We denote the number of LSTM layers of the probabilistic forecast model as $N_L$ and set the size of hidden states for each layer as $H_L$. The forecast model can be represented as $F(\boldsymbol {X}_t\text{, }\alpha\text{; }\theta)$ [as we defined in (\ref{nonp-model})] and we use it to forecast quantiles at the next time step, i.e., $x^{\alpha}_{t+1|t}$. By changing the nominal proportion $\alpha$ in $F(\boldsymbol {X}_t\text{, }\alpha\text{; }\theta)$, different quantiles are forecasted based on $\boldsymbol X_t$ and different hidden states are also generated simultaneously. Here, no a priori hypotheses are required on the probability distribution type, the probabilistic forecast model built is therefore nonparametric.

\subsection {End-to-End Training With Missing Data Imputation}

We organize the historical time series into multiple sequence instances to prepare for model training. The end-to-end training process for the forecast model, considering missing data imputation, is illustrated in Fig.~\ref{end-to-end-trian}. This process primarily involves iterative imputation and model training with iterative loss.

\subsubsection {Iterative Imputation}
\label{recur-impu}

At the top of Fig.~\ref{end-to-end-trian}, the green box indicates that the actual value is observed in the time series at the timestamp marked above it, while the grey box denotes a missing observation. The pink box with a dashed edge represents the corresponding imputed value. A sequence instance $S$ = [$\boldsymbol X_{t}$, $\boldsymbol X_{t+1}$, $\cdots$, $\boldsymbol X_{t+T-1}$] is iteratively generated from the incomplete time series. This sequence records $T$ successive inputs of the model, where each input $\boldsymbol X_{i}$, $i \in$ [$t$, $t+T-1$], is a distributed renewable energy output series covering the lag interval $\delta$ from time spot $i-\delta+1$ to $i$.

Specifically, the first input $\boldsymbol X_{t}$ at time $t$ is denoted as $[x_{t-\delta+1}, \cdots, x_{t}]$, assuming there is no missing value or it has been imputed in $\boldsymbol X_{t}$. The next input $\boldsymbol x_{t+1}$ is formulated as $[x_{t-\delta+2}, \cdots, \hat x_{t+1}]$, with the new information $\hat x_{t+1}$ given by:
\begin{equation}
\hat x_{t+1} = x_{t+1}m_{t+1} + (1-m_{t+1}) x_{t+1|t}^{0.5},
\label{imputation}
\end{equation}
where $m_{t+1}$ is a binary variable indicating the presence (1) or absence (0) of observed data at time spot $t+1$. In cases where data is missing ($m_{t+1}=0$), the missing observation is imputed as $x_{t+1|t}^{0.5} = F(\boldsymbol X_{t}, 0.5; \theta)$, i.e., the median forecasted at the last time spot $t$. Conversely, when data is available ($m_{t+1}=1$), the actual value $x_{t+1}$ is utilized to construct $\boldsymbol X_{t+1}$.

We iteratively implement the imputation process shown in (\ref{imputation}) from time spot $t$ to $t+T-1$ to obtain all the inputs. These inputs are then sent to the forecast model $F_\theta$ (short for $F(\cdot; \theta)$). Hidden states [$h_t$, $h_{t+1}$, $\cdots$, $h_{t+T-1}$] are calculated and passed through the forecast model over $T$ time spots, generating corresponding forecasted quantiles [$x_{t+1|t}$, $x_{t+2|t+1}$, $\cdots$, $x_{t+T|t+T-1}$] via feedforward calculation.


\subsubsection {Model Training With Iterative Loss}

We employ the pinball loss to assess the performance of the probabilistic based on forecasted quantiles and corresponding observations. Considering the forecasted $\alpha$-th quantile $x_{i+1|i}^{\alpha}$ based on $\boldsymbol X_{i}$, the pinball loss is calculated as follows:
\begin{equation}
  L_{x}^i=\left\{
  \begin{array}{l}
    m_{i+1}\sum_{\alpha\in Q}\alpha(x_{i+1}-{x}_{i+1|i}^{\alpha})\text{, }x_{i+1}\geq{x}_{i+1|i}^{\alpha} \\
    m_{i+1}\sum_{\alpha\in Q}(1-\alpha)({x}_{i+1|i}^{\alpha}-x_{i+1})\text{, }x_{i+1}<{x}_{i+1|i}^{\alpha} 
  \end{array}
  \right.
  \end{equation}
where $x_{i+1}$ is the observation of $ x_{i+1|i}^{\alpha}$. Here, we only consider the pinball loss when the actual value is observed at time spot $i+1$, as missing observations are imputed by the forecasted median ($0.5$-th quantile) for time spot $i+1$ when data is missing.

Considering all inputs over $T$ time spots, a iterative loss is formulated through the whole sequence instance, which is defined as
\begin{equation}
  L_{rec}=\frac{1}{T}\sum_{i=t}^{T+t-1} L_{p}^i\text{, }
  \label{rec-loss}
  \end{equation}
and this iterative loss $L_{rec}$ is minimized via gradient descent to optimize parameters $\theta$ of the forecast model $F_{\theta}$. 

Actually, the forecast model is trained using batches of sequence instances to improve training efficiency. The iterative imputation and training processes described above iterate over all batches, constituting a training epoch. We halt the training process when the average training loss has been consistently lower than the validation loss for over 20 successive training epochs, thus preventing overfitting.

\section {Simulations}
\label{sec-5}

\subsection{Data Set and Data Missing Mechanism}

We obtained a real-world wind power dataset from the Australian National Electricity Market \cite{Wind-farm-data}. This dataset contains wind power data for the Bluff Range wind farm located in South Australia, spanning from 2018 to 2020, with a capacity of 52.5 MW. Since the capacity of the wind farm is not large, we take this wind farm as a distributed renewable generation. The wind power data has a time resolution of 5 minutes. The input $\boldsymbol X_t$ for the forecast models includes the historical wind power series. Before forecasting, all features are normalized using min-max normalization. We assume that missingness in the wind power data follows a missing completely at random (MCAR) mechanism, implying that the absence of data is uncorrelated with the values themselves \cite{little2019statistical}. In our experiment, the missing rate is set at 25\%.


\subsection{Benchmarks}

In our comparisons, we consider several state-of-the-art imputation and forecasting benchmarks, including both two-phase and end-to-end methods. Two-phase methods separate the imputation and forecasting into two steps, combining different commonly employed techniques for imputation and forecasting. For imputation, we consider methods linear interpolation (LI) \cite{demirhan2018missing} and k-Nearest Neighbors (KNN) \cite{7444178}. For forecasting, we use LSTM due to its superior performance in renewable forecasting, as validated in \cite{li2022integrated}, and because it shares the same nonparametric structure as our proposed method, ensuring a fair comparison. Specifically, these two-phase benchmarks are referred to as LI-LSTM and KNN-LSTM. For the end-to-end method, we adopt the Autoregressive LSTM (Auto-LSTM) \cite{li2022integrated} as the benchmark, where a single model is used to perform imputation and forecasting simultaneously in a parametric manner. Among these benchmarks, except Rec-LSTM, the rest are nonparametric methods.

\subsection{Model Training Process}

We used 60\% of historical samples for training the model, the middle 20\% for validation, and the last 20\% for testing. If the average loss on the training dataset has been consistently lower than that on the validation dataset over 20 successive epochs, the training process is stopped to prevent overfitting. We used a grid-search method to find the optimal structure of the forecast model (associated with $N_L$ and $H_L$), the lag interval $\delta$, and the learning rate $l_r$. Specifically, $N_L$ was chosen from \{8, 16, 32, 64\}, $H_L$ from \{16, 32, 64, 128\}, $\delta$ from \{5 minutes, 10 minutes, 15 minutes, 20 minutes\}, and $l_r$ from \{1e-4, 1e-3, 1e-2, 1e-1\}. The optimal values for \{$N_L$, $H_L$, $\delta$, $l_r$\} were determined when the average validation loss was the lowest during the grid-search procedure. In our case, this resulted in \{16, 32, 15 minutes, 1e-3\}.

\subsection{Performance Evaluation and Comparison}

We compare the performance of LI-LSTM, KNN-LSTM, and Rec-LSTM under missing rates at 25\%. Average deviation $\overline{R_l}$ in (\ref{a_l}), average PI width $\overline{S_l}$ in (\ref{d_l}), and the average skill score $\overline{Sk_l}$ in (\ref{s_l}) were used to evaluate the probabilistic forecasting results on the testing data set. The average results are recorded in Table~\ref{table_performance}. Illustrations of the results are as follows.

\begin{table}[!tbp]
  \renewcommand{\arraystretch}{1.3}
  \caption{Comparison for probabilistic forecasting for Distributed Energy Resources.
  }
  \label{table_performance}
  \centering
    \resizebox{7cm}{!}{
  \begin{tabular}{c|c|c|c}%
  \hline\hline
    &\makecell*[c]{$\overline{R_l}\%$} &\makecell*[c]{$\overline{S_l}$}  &\makecell*[c]{$\overline{Sk_l}$}\\ \hline
  {LI-LSTM}  
  &$11.4$ &$0.208$ &$-1.22$ \\\hline
  {KNN-LSTM}  
  &$7.13$ &$0.192$ &$-0.974$ \\\hline
  {Auto-LSTM}  
  &$3.48$ &$0.0422$ &$-0.282$ \\\hline
  {Proposed method}  
  &$3.42$ &$0.0415$ &$-0.276$\\
  \hline\hline
    \end{tabular}}
  \end{table}

In the reliability evaluation, our proposed nonparametric end-to-end method provides low deviations and achieves the lowest average deviation $\overline{R_l}$ (3.42\%) shown in Table~\ref{table_performance}, demonstrating the highest reliability. Two-phase benchmarks LI-LSTM and KNN-LSTM show larger deviations in $\overline{R_l}$, indicating lower reliability. Notably, the end-to-end method Rec-LSTM closely rivals our proposal in $\overline{R_l}$ and outperforms other two-phase benchmarks in terms of reliability. In the sharpness evaluation, our proposal showcases the lowest average PI width $\overline{S_l}$ at missing rate 25\%. Conversely, methods like LI-LSTM and KNN-LSTM exhibit larger $\overline{S_l}$. For the comprehensive indicator skill score $\overline{Sk_l}$, the fourth column in Table~\ref{table_performance} confirms the excellence of our proposed nonparametric end-to-end method, with the highest average skill score $\overline{Sk_l}$ ($-$0.276) among all methods, highlighting its superior overall performance. 

In summary, the performance of the end-to-end methods, i.e., our proposal and Rec-LSTM, provide better comprehensive performance than those of the other two-phase methods. Our nonparametric end-to-end method outperforms its parametric counterpart, i.e., Rec-LSTM. Therein, the assumption of Gaussian distribution for probability distributions of nonstationary wind power may not always hold, contributing to the limitations of Rec-LSTM in such scenarios. On the other hand, methods employing deep learning-based imputation techniques, i.e., our proposed method and Rec-LSTM, generally offer better evaluation results compared to those utilizing traditional statistical methods, i.e., LI-LSTM and KNN-LSTM. This can be attributed to the more potent nonlinear approximation capabilities of deep learning techniques, enabling accurate imputation of missing values and subsequently yielding higher performance in probabilistic forecasting of renewable generation's outputs. Our proposed method synergistically combines the advantages of the nonparametric approach, end-to-end structure, and deep learning, resulting in a highly competitive performance considering missing data imputation.

\subsection{Probabilistic Forecasting Results of the Proposed Method}

\begin{figure}[!tbp]
  \begin{center}
  \includegraphics[width=3.5in]{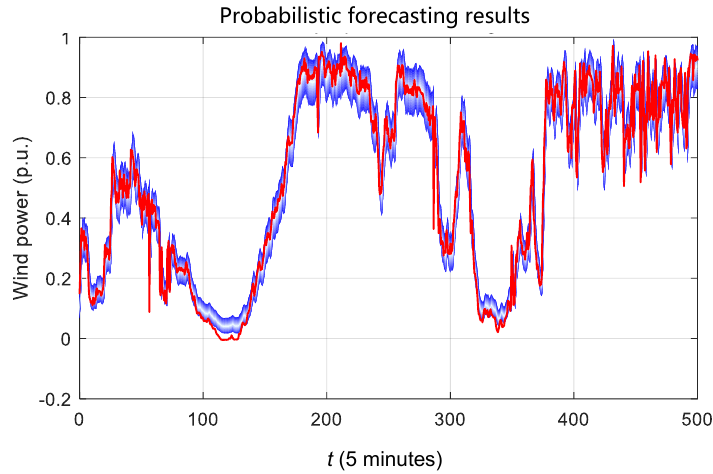}\\
  \caption{
    Probabilistic forecasting of distributed energy resources' outputs.}\label{wpf-results-12}
  \end{center}
\end{figure}

The one-step-ahead probabilistic forecasting results from our proposed method, reflecting confidence levels from 10\% to 90\%, are depicted in Fig.~\ref{wpf-results-12}. Fig.~\ref{wpf-results-12} illustrates forecasting outcomes over 500 successive time spots under scenarios with 25\% missing rate. Fig.~\ref{wpf-results-12}, the red line represents the observation of the forecasting target, and the horizontal axis represents successive time spots with 5-min resolution. One can see that our proposed method consistently provides highly reliable probabilistic forecasting results. The forecasted PIs effectively encompass the observations and accurately capture the up and down trends in the wind power series. The discussions above verify the effectiveness of our proposed method under the missing-data scenario.

\section{Conclusion}

This paper presents a nonparametric end-to-end probabilistic forecasting method for characterizing the stochastic properties of distributed renewable generations while addressing integrated missing data imputation. In our methodology, missing observations are iteratively imputed using the forecasted median. Subsequently, the probabilistic forecast model is trained with a revised loss function based on pinball loss derived from forecasted quantiles, ensuring an end-to-end process. Our proposed methodology exhibits exceptional proficiency in probabilistic forecasting for distributed renewable generations. Through numerical simulations, we convincingly demonstrate the superiority of our method over benchmarks, considering its alignment with real-world data and the width of PI.


\bibliographystyle{IEEEtran}
\bibliography{IEEEabrv,Bibliography}


\end{document}